\documentclass[11pt,twoside]{article}

\usepackage{asp2004}
\usepackage{epsf}
\usepackage{psfig}
\usepackage{lscape}

\newcommand{\oversim}[2]{\protect{\mbox{\lower0.5ex\vbox{%
  \baselineskip=0pt\lineskip=0.2ex
  \ialign{$\mathsurround=0pt #1\hfil##\hfil$\crcr#2\crcr\sim\crcr}}}}}
\newcommand{\simgreat}{\mbox{$\,\mathrel{\mathpalette\oversim>}\,$}} 
\newcommand{\simless} {\mbox{$\,\mathrel{\mathpalette\oversim<}\,$}} 
\begin{document}
\title{The birth and early evolution of star clusters} 
\author{Pavel Kroupa} 
\affil{Argelander Institute for Astronomy, University of Bonn, Auf dem
H\"ugel~71, D-53347 Bonn, Germany}

\begin{abstract} 
Star clusters are observed to form in a highly compact state and with
low star-formation efficiencies, and only 10~per cent of all clusters
appear to survive to middle- and old-dynamical age. If the residual
gas is expelled on a dynamical time the clusters disrupt. Massive
clusters may then feed a hot kinematical stellar component into their
host-galaxy's field population thereby thickening galactic disks, a
process that theories of galaxy formation and evolution need to
accommodate.  If the gas-evacuation time-scale depends on cluster
mass, then a power-law embedded-cluster mass function may transform
within a few dozen Myr to a mass function with a turnover near
$10^5\,M_\odot$, thereby possibly explaining this universal empirical
feature. Discordant empirical evidence on the mass function of star
clusters leads to the insight that the physical processes shaping
early cluster evolution remain an issue of cutting-edge research.
\end{abstract}


\section{Early cluster evolution}
\label{pk_sec:earlyevol}
The star-formation efficiency (sfe), $\epsilon\equiv M_{\rm
ecl}/(M_{\rm ecl} + M_{\rm gas})$, where $M_{\rm ecl}, M_{\rm gas}$
are the mass in freshly formed stars and residual gas, respectively,
is $0.2 \simless \epsilon$ $\simless 0.4$ \citep{LadaLada2003}
implying that the physics dominating the star-formation process on
scales $<10$~pc is stellar feedback.  Within this volume, the
pre-cluster cloud core contracts under self gravity thereby forming
stars ever more vigorously, until feedback energy suffices to halt the
process ({\it feedback-termination}) \citep{WeidnerKroupa2006}.  This
occurs on one to a few crossing times ($\approx 10^6$~yr), and since
each proto-star needs about $10^5$~yr to accumulate about 95~per cent of its
mass \citep{WuchterlTscharnuter2003}, the assumption may be made that
the embedded cluster is mostly virialised at feedback-termination.
Its stellar velocity dispersion, 

\begin{equation}
\sigma \approx \sqrt{G\,M_{\rm
ecl}/(\epsilon \, R)}, 
\end{equation}

\noindent
may then reach $\sigma=40\,$pc/Myr if $M_{\rm ecl} =
10^{5.5}\,M_\odot$ which is the case for $\epsilon\,R < 1$~pc. This is
easily achieved since the radius of one-Myr old clusters is $R\approx
1$~pc with a weak, if any dependence on mass
\citep{Bastian_etal2005}. Very young clusters (age~$\simless 10$~Myr)
would thus appear super-virial, i.e. with a velocity dispersion too
large for the cluster mass.

The above exercise demonstrates that the possibility may be given that
a {\it hot kinematical component} could add to a galactic disk as a
result of clustered star formation for reasonable physical
parameters. Thickened galactic disks may result.  But this depends on

\begin{description}
\item (i) $\epsilon$, 

\item (ii) $R$ (cluster concentration) and 

\item (iii) the ratio
of the gas-expulsion time-scale to the dynamical time of the embedded
cluster, $\tau_{\rm gas}/t_{\rm cross}$.
\end{description}

\subsection{Empirical constraints}
\label{pk_ssec:emp}
The first (i) of these is clearly fulfilled: $\epsilon<40$~per cent
\citep{LadaLada2003}.  The second (ii) also appears to be fulfilled
such that clusters with ages $\simless 1$~Myr have $R\simless1$~pc
independently of their mass. Some well-studied cases are tabulated and
discussed in \cite{Kroupa2005}. Finally, the ratio $\tau_{\rm
gas}/t_{\rm cross}$ (iii) remains uncertain but critical.

The well-observed cases discussed in \cite{Kroupa2005} do indicate
that the removal of most of the residual gas does occur within a
cluster-dynamical time, $\tau_{\rm gas}/t_{\rm cross} \simless
1$. Examples noted are the Orion Nebula Cluster (ONC) and R136 in the
LMC both having significant super-virial velocity dispersions. Other
examples are the Treasure-Chest cluster and the very young
star-bursting clusters in the massively-interacting Antennae galaxy
which appear to have HII regions expanding at velocities such that the
cluster volume may be evacuated within a cluster dynamical time.

A simple calculation of the amount of energy deposited by an O~star
within a cluster crossing time into its surrounding cluster-nebula
also suggests it to be larger than the nebula binding energy
\citep{Kroupa2005}. Furthermore, \cite{BastianGoodwin2006} note that
many young clusters have a radial-density profile signature as
expected if they are expanding rapidly.

Thus, the data suggest the ratio $\tau_{\rm gas}/t_{\rm cross}$ to be
near one, but much more observational work needs to be done to
constrain this number. Measuring the kinematics in very young clusters
would be an extremely important undertaking, because the implications
of $\tau_{\rm gas}/t_{\rm cross}\simless 1$ are dramatic.

To demonstrate these implications it is now assumed that a cluster is
born in a very compact state ($R\approx 1$~pc), with a low sfe
($\epsilon <0.4$) and $\tau_{\rm gas}/t_{\rm cross}\simless 1$. As
noted in \cite{Kroupa2005}, ``in the presence of O~stars, explosive
gas expulsion may drive early cluster evolution independently of
cluster mass''.

\section{Implications}
\label{pk_sec:impl}

\subsection{Cluster evolution and the thickening of galactic disks}
\label{pk_ssec:heat}

As one of the important implications, a cluster in the age range of
$\approx 1-50$~Myr will have an unphysical $M/L$ ratio because it is
out of dynamical equilibrium rather than having an abnormal stellar
IMF \citep{BastianGoodwin2006}.

Another implication would be that a Pleiades-like open cluster would
have been born in a very dense ONC-type configuration and that, as it
evolves, a ``moving-group-I'' is established during the first few
dozen~Myr comprising roughly 2/3rds of the initial stellar population
and expanding outwards with a velocity dispersion which is a function
of the pre-gas-expulsion configuration \citep{Kroupaetal2001}. These
computations were in fact the first to demonstrate, using
high-precision $N$-body modelling, that the re-distribution of energy
within the cluster during the embedded phase and the expansion phase
leads to the formation of a substantial remnant cluster despite the
inclusion of all physical effects that are disadvantageous for this to
happen (explosive gas expulsion, Galactic tidal field and mass loss
from stellar evolution).  

Thus, in this scenario stars form in very compact clusters that have
radii less than about 1~pc and masses larger than a dozen~$M_\odot$,
and in the presence of O~stars the residual gas is removed explosively
leading to loss of stars from the cluster which form a moving~group~I
(an expanding population of sibling stars). A cluster re-forms after
gas expulsion as a result of energy-equipartition during the embedded
phase, or during the expanding phase, and fills its Roche lobe in the
Galactic tidal field. Such a cluster appears with an expanded core
radius (compare the ONC with about 0.2~pc and the Pleiades with
1.3~pc) and evolves secularly through evaporation until it's demise
\citep{BaumgardtMakino2003,Lamers_etal2005a,Lamers_etal2005b}.  A
``moving-group-II'' establishes during this stage as the ``classical''
moving group made-up of stars which slowly diffuse/evaporate out of
the re-virialised cluster remnant with relative kinetic energy close
to zero. Under unfavorable conditions no cluster re-forms after gas
expulsion.  Such conditions may arise from time-variable tidal fields
(e.g. through nearby cloud formation) or simply through unfavorable
density profiles just before gas-expulsion
\citep{BoilyKroupa2003a,BoilyKroupa2003b}, and they may dominate over
the favorable ones (\S~\ref{pk_ssec:mass_indep}).  Note that the
number of stars in moving-group-I always outnumber the number of
stars in moving-group-II, while the unfavourable conditions would
lead to no moving-group-II.

Moving-groups-I would be populated by stars that carry the initial
kinematical state of the birth configuration into the field of a
galaxy.  Each generation of star clusters would, according to this
picture, produce overlapping moving-groups-I (and~II), and the overall
velocity dispersion of the new field population can be estimated by
adding in quadrature all expanding populations. This involves an
integral over the embedded-cluster mass function, $\xi_{\rm
ecl}(M_{\rm ecl})$, which describes the distribution of the stellar
mass content of clusters when they are born \citep{Kroupa2002,
Kroupa2005}. It is known to be a power-law
\citep{LadaLada2003,Hunter_etal2003,ZhangFall1999}.  The integral can
be calculated for a first estimate of the effect.  The result is that
for reasonable upper cluster mass limits in the integral, $M_{\rm
ecl}\simless10^5\,M_\odot$, the observed age--velocity dispersion
relation of Galactic field stars in the solar neighbourhood can be
reproduced.

This theory can thus explain the ``energy deficit'', namely that the
observed kinematical heating of field stars with age cannot, until
now, be explained by the diffusion of orbits in the Galactic disk as a
result of scattering on molecular clouds, spiral arms and the bar
\citep{Jenkins1992}. Because the age--velocity-dispersion relation for
Galactic field stars increases with stellar age, this notion can also
be used to map the star-formation history of the Milky-Way disk by
resorting to the observed correlation between the star-formation rate
in a galaxy and the maximum star-cluster mass born in the population
of young clusters \citep{Weidneretal2004}.

A possible cosmologically-relevant implication of this
``popping-cluster'' mo\-del emerges as follows
\citep{Kroupa2002,Kroupa2005}: A thin galactic disk which experiences a
significant star burst such that star clusters with masses ranging up
to $10^6\,M_\odot$ can form in the disk, will puff-up and obtain a
thick-disk made-up of the fast-moving velocity wings of the expanding
clusters after residual gas expulsion. A notable example where this
process may have lead to thick-disk formation is the galaxy UGC~1281
\citep{Mould2005}, which, by virtue of its LSB character, could not
have obtained its thick disk as a result of a merger with another
galaxy.  \cite{ElmegreenElmegreen2006} observe edge-on disk galaxies
with a chain-like morphology in the HST ultra-deep field finding many
to have thick disks. These appear to be derived from kpc-sized clumps
of stars. These clumps are probably massive star-cluster complexes
which form in perturbed or intrinsically unstable gas-rich and young
galactic disks.  The individual clusters in such complexes would
``pop'' as in the theory above \citep{FellhauerKroupa2005}, leading
naturally to thick disks as a result of vigorous early star formation.

\subsection{Structuring the initial cluster mass function}
\label{pk_ssec:mfn}

Another potentially important implication from this theory of the
evolution of young clusters is that {\it if} the gas-expulsion
time-scale and/or the sfe varies with initial (embedded) cluster mass,
then an initially featureless power-law mass function of embedded
clusters will rapidly evolve to one with peaks, dips and turnovers at
``initial'' cluster masses that characterize changes in the broad
physics involved, such as the gas-evacuation time-scale. 

{\it Note} that the ``embedded'' cluster mass is the birth mass in stellar
content, while the ``initial'' cluster mass is the mass of stars the
virialised post-gas-expulsion star cluster would have had if it had
been born with a SFE of 100~per cent ($\epsilon = 1$).

``Initial'' cluster masses are
derived by dynamically evolving observed clusters backwards in time to
$t=0$ by taking into account stellar evolution and a galactic tidal
field but not the birth process \citep[``classical'' evolution tracks,
eg.][]{Portetal2001, BaumgardtMakino2003}.

To further quantify this issue, \cite{KroupaBoily2002} assumed that the function

\begin{equation}
M_{\rm icl} = f_{\rm st}\,M_{\rm ecl},
\end{equation}

\noindent
exists, where $M_{\rm ecl}$ is as above, $M_{\rm icl}$ is the
``classical initial cluster mass'' and

\begin{equation}
f_{\rm st} = f_{\rm st}(M_{\rm ecl}). 
\end{equation}

\noindent
Thus, for example, for the Pleiades, $M_{\rm cl}\approx 1000\,M_\odot$
at the present time (age: about 100~Myr), and a classical initial
model would place the initial cluster mass at $M_{\rm icl}\approx
1500\,M_\odot$ by using standard $N$-body calculations to quantify the
secular evaporation of stars from an initially bound and virialised
``classical'' cluster \citep{Portetal2001}. If, however, the sfe was
33~per cent and the gas-expulsion time-scale was comparable or shorter
than the cluster dynamical time, then the Pleiades would have been
born in a compact configuration resembling the ONC and with a mass of
embedded stars of $M_{\rm ecl}\approx 4000\,M_\odot$
\citep{Kroupaetal2001}.  Thus, $f_{\rm st}(4000\,M_\odot) = 0.38$ in
this particular case.

By postulating that there exist three basic types of embedded
clusters, namely 

\begin{itemize}

\item clusters without O~stars (type~I: $M_{\rm ecl}\simless
10^{2.5}\,M_\odot$, e.g.  Taurus-Auriga pre-main sequence stellar
groups, $\rho$~Oph),

\item
clusters with a few O~stars (type~II: $10^{2.5} \simless M_{\rm
ecl}/M_\odot \simless 10^{5.5}$, e.g. the ONC), and

\item
clusters with many O~stars and with a velocity dispersion
comparable to the sound velocity of ionized gas (type~III: $M_{\rm
ecl}\simgreat 10^{5.5}\,M_\odot$),

\end{itemize}

\noindent
it can be argued that $f_{\rm st}\approx 0.5$ for type~I, $f_{\rm
st}<0.5$ for type~II and $f_{\rm st}\approx 0.5$ for type~III. 

The reason for the high $f_{\rm st}$ values for types~I and~III is
that gas expulsion from these clusters may be longer than the cluster
dynamical time because there is no sufficient ionizing radiation for
type~I clusters, or the potential well is too deep for the ionized gas
to leave (type~III clusters). Type~I clusters are excavated through
the cumulative effects of stellar radiation and outflows, while
type~II clusters probably require multiple supernovae of type~II to
unbind the gas \citep{Goodwin1997}.  According to the present notion,
type~II clusters undergo a disruptive evolution and witness a high
``infant mortality rate'' \citep{LadaLada2003}, therewith being the
pre-cursors of OB associations and open Galactic clusters.

Under these conditions and an assumed inverted Gaussian functional
form for $f_{\rm st}=f_{\rm st}(M_{\rm ecl})$, the power-law embedded
cluster mass function transforms into a cluster mass function with a
turnover near $10^5\,M_\odot$ and a sharp peak near $10^3\,M_\odot$
\citep{KroupaBoily2002}. This form is strongly reminiscent of the
initial globular cluster mass function which is inferred by
e.g. \cite{Vesperini1998,Vesperini2001,ParmentierGilmore2005,BaumgardtMakino2003}
to be required for a match with the evolved cluster mass function that
is seen to have a universal turnover near $10^5\,M_\odot$.  Thus,
$\approx 10^{10}$~yr-old cluster populations have log-normal Gaussian
mass functions quite independently of the environment.  

In the theory presented here, the power-law embedded cluster mass
function evolves in the first~$10^{7-8}$~yr to the ``initial'' cluster
mass function with the broad turn-over near $10^5\,M_\odot$. In this
context it is of interest to note that \cite{deGrijs_etal2003} found a
turnover in the cluster mass function near $10^5\,M_\odot$ for
clusters  with an age near $10^9$~yr in the star-burst galaxy M82. This
result was challenged, but in their section~2.1 \cite{deGrijs_etal2005}
counter the criticisms.  This age of 1~Gyr is too young to be
explainable through secular evolution starting from a power-law
cluster mass function.  In contrast to this interpretation, the theory
outlined here would allow a power-law embedded cluster mass function
to evolve to the observed form within 1~Gyr. Evaporation through
two-body relaxation in a time-variable tidal field then takes over
thereby enhancing the observed log-normal shape of the old cluster
mass function
\citep[eg.]{Vesperini1998,Vesperini2001,ParmentierGilmore2005}.

This ansatz may thus bear the solution to the long-standing problem
that the initial cluster mass function needs to have this
turnover, while the observed mass functions of very young clusters are
featureless power-law distributions.

\subsection{Cluster disruption independently of mass}
\label{pk_ssec:mass_indep}

If, on the other hand, $f_{\rm st} =\,$constant such that only about
10~per cent of all clusters survive independently of mass, then the
above long-standing problem would remain unsolved.  This situation is
suggested by the results of \cite{LadaLada2003} and
\cite{Fall_etal2005}, among others.  Observations show that 90~per
cent of all very young clusters dissolve within about $10^4$~yr, independently of
their mass. This may either be due to a constant $f_{\rm st}$, or, as
pointed out by \cite{Fall_etal2005}, a result of 90~per cent of all
clusters dissolving and only the rest surviving intact. The former
case would simply imply a constant shift of the power-law mass
function along the mass axis to smaller masses as a result of cluster
disruption through stellar feedback (operating in all embedded clusters), while
the latter would imply a purely stochastic element to the disruption
of clusters. 

It would therefore appear that this 'high infant mortality of clusters 
independently of cluster mass' scenario would be in contradiction with  
the results of \cite{deGrijs_etal2005} on the M82 1~Gyr-old cluster system
combined with the observed power-law mass function of very young clusters. 

\section{Conclusions}
\label{pk_sec:concs}

Observations show that star clusters of any mass are formed extremely
compact ($\simless 1$~pc), while older clusters appear well distended
with radii of a few-to-many~pc. Very young clusters seem to be
super-virial, and expanding HII shells also indicate the explosive
removal of residual gas.  Star-cluster formation thus appears to be
rather violent leading to a loss of a large fraction of stars and
probably in many cases total cluster disruption.

As this discussion shows, the formation process of star clusters may
have significant cosmological implications in that the morphology of
galaxies may be shaped by the clusters born within them: the disks of
galaxies are probably thickened during periods of high star-formation
rates while low-mass dwarf galaxies may attain halos of stars as a
result of the violent processes associated with cluster birth as a
result of which stars may spread outwards with relatively high
velocity dispersions.  Particular examples of such events have been
noted above.  If this is true can be verified by measuring the
velocity dispersions of stars in clusters younger than at most a
few~Myr. This is very important to do because thickened galactic disks
are often taken to be evidence for the merging of cold-dark-matter
sub-structures. Naturally, in-falling dwarf galaxies will lead to
thickened disks, and counter-rotating thick disks would be prime
examples of such processes \citep{YoachimDalcanton2005}, but this
discourse shows that sub-pc-scale processes would probably need to be
incorporated in any galaxy-evolution study.

The mass function of old globular clusters has a near-universal
turnover near $10^5\,M_\odot$ which has so far defied an explanation,
given that the mass function of very young clusters is observed to be
a power-law, and that the modelling of classical disruption through
two-body relaxation and tidal fields poses a challenge for evolving a
power-law to the approximate log-normal form for old clusters.  Again,
mass-dependent processes related to the physics of residual gas
expulsion can explain such a turnover, but in this case the turnover
would have to become evident in cluster populations younger than about
100~Myr.  The results of \cite{deGrijs_etal2005} suggest this to be
the case.  The results of \cite{ZhangFall1999}, however, appear to
show no such evidence for the cluster population in the Antennae
galaxies: the mass function for two age groups of clusters (younger
than about 10~Myr, and between 10 and 160~Myr) show the same power-law
form, albeit with different normalisations as a result of the high
infant mortality rate. If this is true, then orbital anisotropies in
the young globular cluster populations as proposed by
\cite{FallZhang2001} would be needed to evolve a power-law mass
function to the log-normal form.

It therefore appears that the broad principles of early cluster
evolution remain quiet unclear.  Given that {\it clusters are the
fundamental building blocks of galaxies} \citep{Kroupa2005} this is no
satisfying state of affairs.  On the positive side, the importance of
star clusters to stellar populations and other cosmological issues
(merging histories of galaxies, galactic morphology) means that much
more theoretical and observational effort is needed to clarify which
processes act during the early life of a star cluster, and how these
affect its appearance at different times.

\acknowledgements I thank the organisers for giving me an opportunity
to present this material at this great meeting which was
characterised by a wonderful atmosphere of friendship towards Henny Lamers
by all participants. 


\end{document}